\documentstyle[prl,aps,floats]{revtex}
\input{psfig.tex}

\tighten
\begin{document} 
\draft
\twocolumn[\hsize\textwidth\columnwidth\hsize\csname@twocolumnfalse\endcsname
\preprint{}
\title{Structure formation by cosmic strings with a cosmological constant}
\author{Richard A. Battye{$^{1}$}, James Robinson{$^{2}$} and 
Andreas Albrecht{$^{3}$}}
\address{
${}^1$ Department of Applied Mathematics and Theoretical Physics, University
of Cambridge, \\ Silver Street, Cambridge CB3 9EW, U.K. \\
${}^2$ Department of Astronomy, Campbell Hall, University of California,
Berkeley CA 94720, U.S.A.\\
${}^3$ Theoretical Physics Group, Blackett Laboratory, Imperial
College, Prince Consort Road, 
 London SW7 2BZ,  U.K.
}
\maketitle
\begin{abstract}
%\narrowtext
We investigate cosmic string models for structure formation which
include a non-zero cosmological constant. We find that the background
evolution of density perturbations and modifications to the  scaling
behaviour of the strings both act to increase the amount of power present
on $100h^{-1}$Mpc scales. We estimate the size of this effect using an
analytic model for the evolution of the string network and find that a
bias $b\sim 2$ can give acceptable agreement with the current
observations. An interesting by product of these modifications is a
broad peak in the cosmic microwave background (CMB) angular power spectrum
around $l=400-600$. 
\end{abstract}
\date{\today}

]
\renewcommand{\thefootnote}{\arabic{footnote}}
\setcounter{footnote}{0}

Topological defect theories, such as cosmic strings \cite{kib1}, have
for many years represented the only serious alternatives to the
standard inflationary paradigm as theories for structure formation.  
In inflationary theories, structure in the universe is seeded by
causal processes operating outside the framework of the Standard Big
Bang (SBB) cosmology, which imprint a primordial spectrum of adiabatic
density perturbations on all scales. 
By contrast, in defect theories the universe is taken to be initially
smooth and perturbations must be seeded by the causal action of an
evolving network of defect seeds, within the background of the SSB cosmology.

Recent work \cite{PSelTa,ABRa,ABRb} has pointed to a serious problem for
defect theories reconciling the amplitude of large scale CMB
anisotropies with that in the matter distribution on scales
around $100h^{-1}$Mpc, in a flat universe where the matter density is
critical, that is $\Omega_m=1$. In refs.\cite{ABRa,ABRb}, this problem was
quantified in
terms of $b_{100}$, where $b_R$ is the bias between the
cold dark matter (CDM) and the galaxy distribution in a sphere of
radius $Rh^{-1}$Mpc that we must postulate in order to make a given
theory consistent with the data. It was argued  that unacceptably
large biases,
$b_{100}\simeq 5$, would be required in order to make scaling defect
theories with  $\Omega_{m}=1$ compatible with current observations.
We named this
the $b_{100}$ problem and suggested that its resolution is the most
serious challenge facing defect models for structure formation.  

These recent calculations (see also ref.\cite{ACDKSS}) represent the
first calculations using a linear Einstein-Boltzmann
solver\footnote{The present calculations and those of
refs.\cite{PSelTa,ABRa,ABRb} used a modified version CMBFAST
\cite{cmbfast}  
which has been shown to give accurate $(\sim 1\%)$ results for most
inflationary models.} which includes all the relevant physical
effects. The only input required is the unequal time-correlator (UETC)
for the defect stress-energy, or an ensemble of source histories with
the same UETC. In refs.\cite{ABRa,ABRb} we used a sophisticated model
to represent the stress-energy of a network of cosmic strings, while
numerical simulations were used in refs.\cite{PSelTa,ACDKSS}.  

In the past, uncertainties in the results of defect calculations have meant
that work has tended to concentrate on the simplest possible models. 
Some
attempts have been  made to modify the semi-analytic approach of
Albrecht and Stebbins \cite{AS}
to include additional parameters,
such as  curvature (an open universe, $\Omega_k$) or a cosmological
constant $(\Omega_{\Lambda}$)\cite{Fer,ACM}. 
This method yields only the shape of the matter power spectrum,
although in ref.\cite{ACM} a simple model was employed to estimate the
dependence of the large angle fluctuation amplitude on $\Omega_\Lambda$ and
$\Omega_k$, allowing rough statements about the relative CMB/matter
normalization to be made. The results of these
calculations --- reduced bias required to reconcile theory with the
data, and an improvement to the shape of the matter spectrum
 --- serve as impetus for the current work, which considers a flat
universe with a non-vanishing cosmological constant. Our calculations
use the same modified version of CMBFAST and source two-point
functions as in refs.\cite{ABRa,ABRb} and  represent a self consistent
calculation of both the CDM and CMB power spectra over all observable
scales. We should note that similar results for the matter power
spectra have been found by Avelino {\it et al} \cite{ASW}. 

\begin{figure}[t]
\centerline{\psfig{file=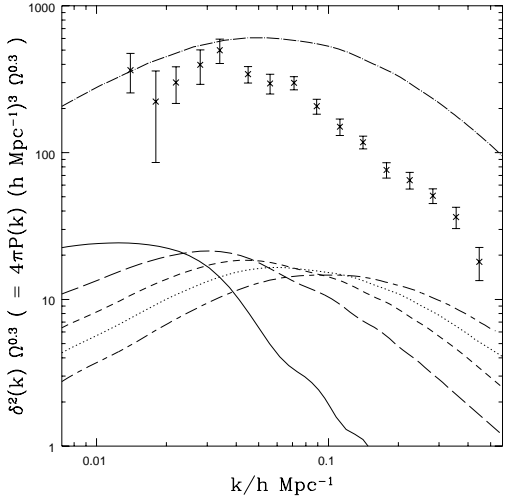,width=2.75in}}
\caption{The (COBE normalized) matter power 
from the scaling cosmic string model for various values
$\Omega_\Lambda$. $\Omega_{\Lambda}=0.0$
(short-long dashed line), 0.3 (dotted dashed line), 0.5 (short dashed
line), 0.7 (long dashed line) and 0.9 (solid line). Included also are
the observed  data points \protect\cite{PD} and the Standard Cold
Dark Matter model (dot-dashed curve). } 
\label{back_mat}
\end{figure}

The introduction of a cosmological constant will be seen to improve
the required value of the bias in two ways. The first is in terms of
the modified evolution of the background spacetime, which is illustrated in
Fig.~\ref{back_mat}. This shows the results for our standard scaling
string source (described in detail in 
\cite{ABRb}) with various choices of the parameter $\Omega_\Lambda$, the
proportion of the critical density from the cosmological constant with the
Hubble constant $H_0=100h\,{\rm km}\,{\rm sec}^{-1}\,{\rm Mpc}^{-1}$ given
by $h=0.5$. We find that as $\Omega_\Lambda$ is increased, the amount of
power on large scales (around
$100h^{-1}{\rm Mpc}$) is increased, with a corresponding decrease on small
scales,
bringing the shape of the matter spectrum in line with the galaxy
data and reducing the magnitude of the $b_{100}$ problem. This
modification can be understood via the standard shape parameter
$\Gamma\approx\Omega_{m}h$ and the equivalent shift in the time of
radiation-matter equality. We find that $\Omega_{\Lambda}=0.7$ seems
to reproduce the shape of the observed spectrum particularly well,
although assuming perfect scaling this still requires a
bias $b\approx 4$, over all scales. Note that we plot
$P(k)\Omega_{m}^{0.3}$ rather than just $P(k)$, so as to undo the
redshift space distortion calculated in ref.\cite{PD}.  

\begin{figure}[t]
\centerline{\psfig{file=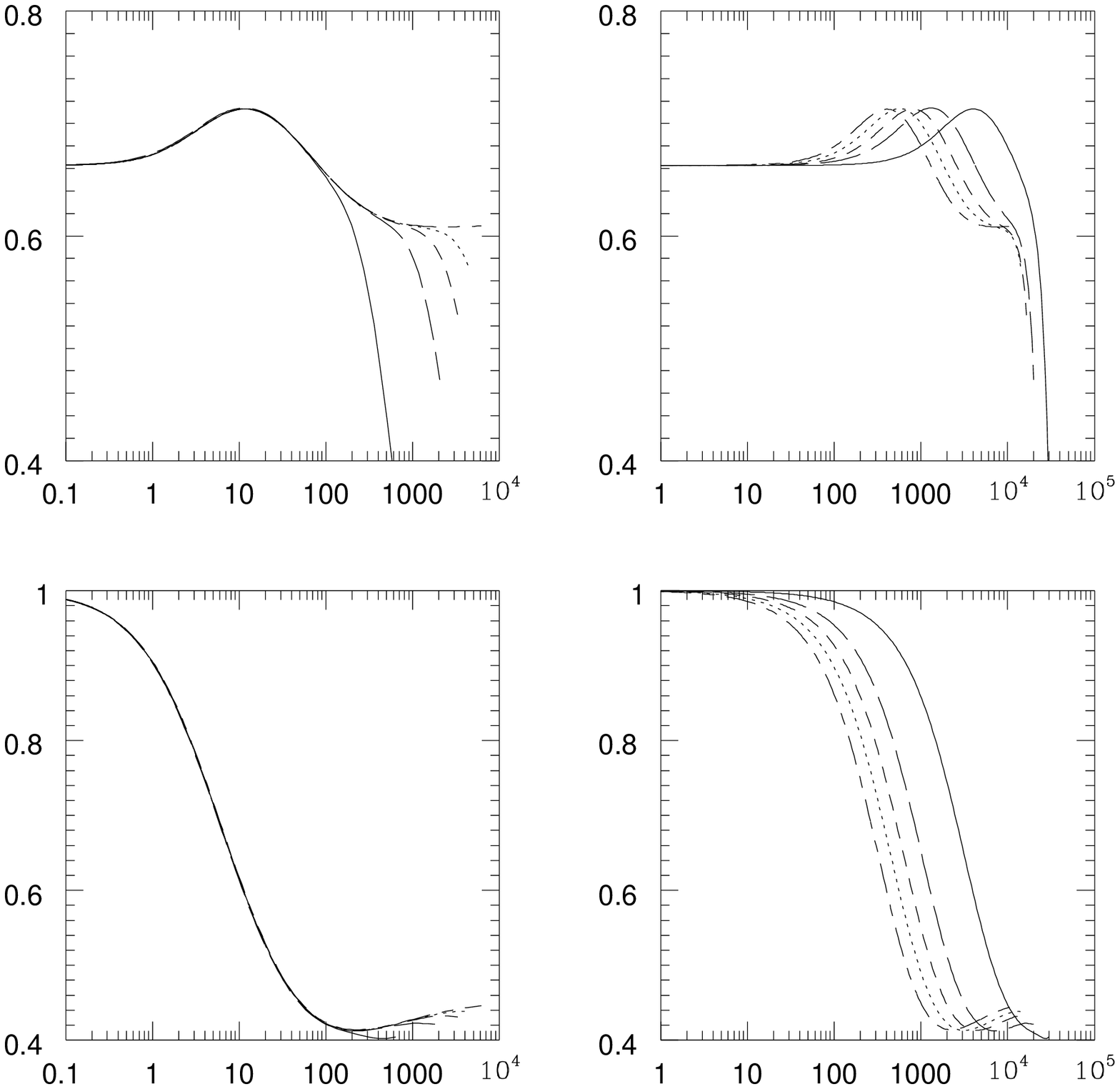,width=2.75in}}
\caption{The results of the velocity dependent one-scale model for the same
values of $\Omega_{\Lambda}$ as in Fig.~\ref{back_mat}. On left the density
of strings relative to the background against $a/a_{\rm eq}$ (bottom) and
the rms velocity of the strin
gs (top). On the right the same quantities but this time against $\tau$.} 
\label{scaling}
\end{figure}

\begin{figure}[t]
\centerline{\psfig{file=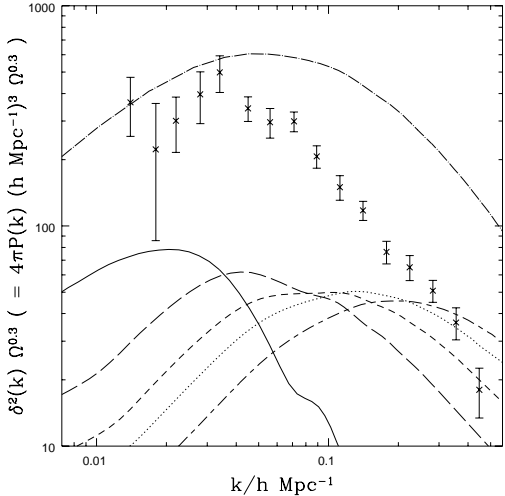,width=2.75in}}
\caption{The (COBE) normalized matter power spectrum for the same values of
$\Omega_{\Lambda}$ as in Fig.~\ref{back_mat}, but this time using the
velocity dependent one scale model to calculate $\xi$ and $v$. Once again
we include the observed data point
s and Standard CDM.}
\label{vel_mat}
\end{figure}

\begin{figure}[t]
\centerline{\psfig{file=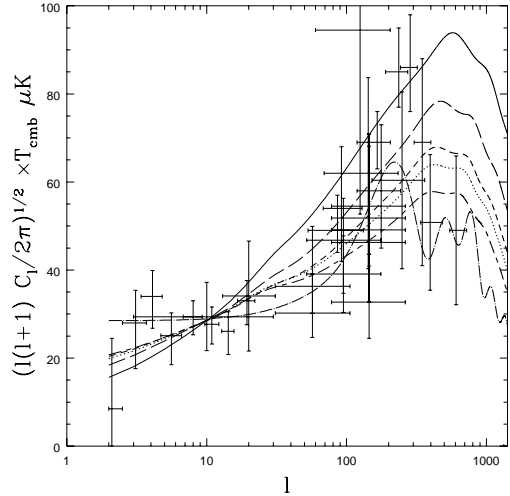,width=2.75in}}
\caption{The CMB power angular power spectrum for the same values of
$\Omega_{\Lambda}$ as in Fig.~\ref{back_mat}. Also included are the
observations data points \protect\cite{tegmark} and the standard CDM curve,
(dot-dashed) curve for comparison.}
\label{vel_cl}
\end{figure}

In our previous work\cite{ABRa,ABRb} we have made an extensive study
of the effects of deviations from scaling. In the case of
$\Omega_{\Lambda}=0.0$, we found that only very radical deviations
from scaling can improve the amount of power on $100h^{-1}$Mpc scales,
at the expense of creating an over production of power on smaller
scales. However, given the improvement in the shape and amplitude of
the matter power spectrum from the inclusion of a cosmological
constant, one might think that mild deviations from scaling might
improve things further.  

Rather than using the  arbitrary scaling deviations to deduce a
transition which yields the best case scenario, for the bulk of this
work we have used the velocity dependent one-scale model for string
evolution \cite{MSa}, which was also used as the basis for the string
evolution in ref. \cite{ACM}. This allows us to calculate the three quantities
which are variables in our model, namely the string coherence length
relative to the horizon $\xi(\tau)$, the rms string velocity $v(\tau)$
and the string density $\rho_s(\tau)=\mu/ \xi^{2}$, as functions of
conformal time, where $\mu$ is the mass per unit length of the
string. The evolution equations for this model become 
\begin{equation} 
\ell^{\prime}={\cal H}\ell v^2+{1\over 2}{\tilde c}v\,,\quad
v^{\prime}=(1-v^2)\left({{\tilde k}\over \ell} - 2{\cal H}v\right)\,,
\end{equation}
where $\ell$ is the length scale on the network, ${\cal H}=a^{\prime}/a$,
${\tilde c}(\tau)$ is the chopping efficiency and ${\tilde k}(\tau)$ is the
curvature of the strings. In order to interpolate between the measured
values in the radiation and matter
 dominated eras, we use 
\begin{equation}
{\tilde c}(\tau)={c_{\rm r}+a(\tau)gc_{\rm m}\over 1+a(\tau)g}\,,\qquad
{\tilde k}(\tau)={k_{\rm r}+a(\tau)gk_{\rm m}\over 1+a(\tau)g}\,,
\end{equation}
where  $c_{\rm r}=0.23,k_{\rm r}=0.17$ and $c_{\rm m}=0.18,k_{\rm m}=0.49$,
the values in the radiation and matter eras respectively, and $g=0.19$ is a
constant which determines the speed of the transition. It has been suggested
\cite{MSb} that this model
 reproduces the behaviour of a string network through the radiation-matter
transition.

The solution of these equations is illustrated in Fig.~\ref{scaling}
for various values of $\Omega_{\Lambda}$ and $h=0.5$. We see that the
density of strings shifts by approximately a factor of two in the
interval $a/a_{\rm eq}$=1 to 100 independent of the value of
$\Omega_{\Lambda}$ and in  the models with $\Omega_{\Lambda}\ne
0$ there is a further shift in the string density, during the
$\Lambda$ dominated phase; the effect being most obvious for large
$\Omega_{\Lambda}$. But when plotted against $\tau$, we see that
$\tau_{\rm eq}$, and hence the transition era, is shifted as $\Omega_m^{-1}$
in conformal time. For
$\Omega_m=1$, the velocity interpolates between the measured values in
the radiation and matter eras, with a transient increase during  the
transition, and for $\Omega_{\Lambda}\ne 0$, the velocity of the
strings is exponentially suppressed during the $\Lambda$ dominated
era. Physically, the cosmological constant is causing the universe to
expand exponentially, reducing the velocity of the strings, preventing
reconnection and hence the strings are no longer scaling.  

Here, we comment briefly on the implementation of these deviations from
scaling into our string model. The velocity and length of each string
segment at each time are simply chosen to be $v(\tau)$ and
$\xi(\tau)$, but for technical reasons we now pick a fixed
value of $v$ rather than choosing it from a random distribution, which
has been shown to make very little difference to the resulting power
spectrum. The number density of strings is taken to be
$(\xi(\tau)\tau)^{-3}$, that is, there is one string per correlation
length cubed. 

Figs.~\ref{vel_mat} and ~\ref{vel_cl} show the 
matter power spectra and CMB anisotropies for the same values of
$\Omega_{\Lambda}$ as Fig.~\ref{back_mat}, with $\xi$, $v$ and number
density of strings given by this model. 
Coupled with the effects of the background, the deviations from
scaling improve $b_{100}$ significantly for $\Omega_{\Lambda}\ne 0$,
with the best fit for $\Omega_{\Lambda}=0.7$ (and hence, for a similar
value of $\Omega_m$ as favoured in refs.\cite{Fer,ACM}), although the
inclusion of the time dependence on $\xi$ has slightly affected the
shape of the power spectrum. We estimate that a bias of $b_{100}=2.6$
would give a 
good fit to the observations on large scales at the expense of an
overproduction of power on smaller scales. A robust feature
of the CMB spectra appears to be a peak at around $l=400-600$, whose
height is increased as $\Omega_{\Lambda}$ increases. Finally, we
should note that the value of string density per unit length  $G\mu$
required to normalize to COBE is around $1\times 10^{-6}$, almost
independent of $\Omega_{\Lambda}$, if we assume that the effective
mass per unit length is $\tilde\mu\approx 1.7\mu$. This is well below
the upper bound imposed by the absence of timing residuals in
measurements of milli-second pulsars \cite{CBS}.  

\begin{figure}[t]
\centerline{\psfig{file=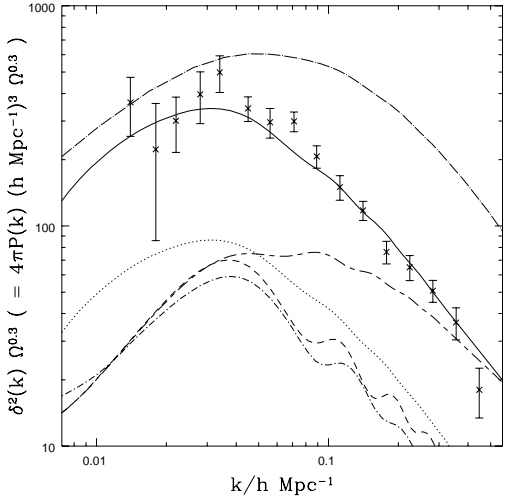,width=2.75in}}
\caption{The matter power spectrum for various models with
$\Omega_\Lambda=0.7$. Velocity one scale model: (1) $h=0.7$, $\Omega_b=0.05$
-- long-short dash line;
(2) $h=0.7$, $\Omega_b=0.15$ -- short dash line; 
(3) $h=0.5$, $\Omega_b=0.125$ -- dot-short dash line. Transition model
with $\chi=4$: unbiased -- dotted curve; biased -- solid curve.} 
\label{fig-best}
\end{figure}

We have shown, therefore, that the inclusion of a cosmological
constant can improve the amount of matter 
power of larger scales, and that an interesting by product of this is a peak in
the CMB angular power spectrum. We now illustrate that once the lack
of large scale power has been remedied, it is relatively simple
improve the fit to the observations on smaller scales. Ignoring for
the moment the constraints on $\Omega_bh^2$
from Big Bang Nucleosynthesis (BBN), 
Fig.~\ref{fig-best} contains three curves
created using the velocity dependent one-scale model and
$\Omega_{\Lambda}=0.7$ with (1) $\Omega_b=0.05$ and $h=0.7$, (2)
$\Omega_b=0.15$ and $h=0.7$, (3) $\Omega_b=0.125$ and $h=0.5$. Model
(1) has the distinction that it fits the amplitude of the small scale
matter data ($b_8\sim 1.0$) though the shape is clearly not right,
with a bias of 2.4 on 100$h^{-1}$Mpc scales. Models (2) and (3) rectify
the small scale slope 
of the spectrum by increasing the baryon fraction, which damps out small
scale power. A similar effect could be
obtained by the introduction of hot dark matter.

As one final point we return to the approach adopted in
refs. \cite{ABRa,ABRb} of
finding the transition which fits the data best, since the exact 
nature of the deviation from scaling at the onset of cosmological
constant domination has not yet been simulated. We implement a
transition using the model described in ref. \cite{ABRb} by varying
$\mu$ with $\tau_T=10000$, $L_T=0.1$ and $\chi=4$, such that the time and
length of the transition closely mirror those obtained in the
one-scale velocity model  and the amplitude chosen to give the best
fit to the data. The results are shown
in Fig.~\ref{fig-best} first without bias and, to emphasize the goodness of
the fit to the shape of the matter data, with a scale invariant bias of
$b=2.0$. This model has been shown by Gawiser and Silk \cite{GS} to
give a good fit to the entire dataset, including CMB and matter power
spectra, peculiar velocities and cluster abundances.

Our results demonstrate that some of the problems of standard defect
scenarios in underproducing power in the large scale matter
distribution can be remedied  by the introduction of a cosmological
constant. Within this scenario,  there are two distinct effects, a
change  in the
evolution of the background perturbations, and more importantly
a modification to the
scaling behaviour of 
the strings, which both act to allow
more matter power on scales of 100$h^{-1}$Mpc. Using the velocity dependent
one scale model we find that biases $b\sim 2$ are required: On first
examination, this value still seems unreasonably high\cite{GS}, but
the discrepancy with the data is small enough that these models 
deserve further consideration. 
The improvement to the bias does not contradict the findings of
refs.\cite{ABRa,ABRb}, since it has come about by significant deviation in
the scaling behaviour of the strings, due to the presence of the
cosmological constant. Our
results for the matter power spectrum and bias are in good agreement
with ref. \cite{ACM}, lending weight to the simpler techniques
employed in that work. It is
clear that similar results will also be obtained in an open universe
\cite{ACM,ASW}, and we are currently investigating this case.

Finally, we should comment on our choice of cosmological parameters
since this is a subject which is currently evolving rapidly. It
appears that the best fit to the power spectrum is attained by using
$\Omega_{\Lambda}=0.7$ which is very much in line with 
 the current measurements of cosmological deceleration from supernova
 type Ia \cite{sn1a} and other independent measurements of the matter
 density of the universe.  We should also note  
that some of the values of $\Omega_{\rm b}$ used in
 Fig.~\ref{fig-best} are slightly larger than usually used on the
 basis of BBN. However, these higher values are not excluded
 \cite{olive} and we only use them here to illustrate that models which
 have a reasonable $b_{100}$ can be made to fit the observed power
 spectrum on smaller scales --- for example, the addition of hot dark
 matter can have a similar effect. We conclude that the interaction
 between this model and observations in the coming months will be  
interesting, and could in the end produce strong support for our model.

We thank U. Seljak and M. Zaldariagga for the use of CMBFAST. We would
also like to thank P. Ferreira, N.Turok, E. Gawiser, P. Viana for
helpful comments  and P. Shellard for enlightening discussions
concerning deviations from scaling. This work was supported by PPARC
and computations  
were done at the UK National Cosmology Supercomputing Center, supported by
PPARC, HEFCE and Silicon Graphics/Cray Research. RAB is funded by Trinity
College.

\def\jnl#1#2#3#4#5#6{\hang{#1, {\it #4\/} {\bf #5}, #6 (#2).}}
\def\jnltwo#1#2#3#4#5#6#7#8{\hang{#1, {\it #4\/} {\bf #5}, #6; {\it
ibid} {\bf #7} #8 (#2).}} 
\def\prep#1#2#3#4{\hang{#1, #4.}} 
\def\proc#1#2#3#4#5#6{{#1 [#2], in {\it #4\/}, #5, eds.\ (#6).}}
\def\book#1#2#3#4{\hang{#1, {\it #3\/} (#4, #2).}}
\def\jnlerr#1#2#3#4#5#6#7#8{\hang{#1 [#2], {\it #4\/} {\bf #5}, #6.
{Erratum:} {\it #4\/} {\bf #7}, #8.}}
\def\prl{Phys.\ Rev.\ Lett.}
\def\pr{Phys.\ Rev.}
\def\pl{Phys.\ Lett.}
\def\np{Nucl.\ Phys.}
\def\prp{Phys.\ Rep.}
\def\rmp{Rev.\ Mod.\ Phys.}
\def\cmp{Comm.\ Math.\ Phys.}
\def\mpl{Mod.\ Phys.\ Lett.}
\def\apj{Ap.\ J.}
\def\apjl{Ap.\ J.\ Lett.}
\def\aap{Astron.\ Ap.}
\def\cqg{Class.\ Quant.\ Grav.} 
\def\grg{Gen.\ Rel.\ Grav.}
\def\mn{MNRAS}
\def\ptp{Prog.\ Theor.\ Phys.}
\def\jetp{Sov.\ Phys.\ JETP}
\def\jetpl{JETP Lett.}
\def\jmp{J.\ Math.\ Phys.}
\def\zpc{Z.\ Phys.\ C}
\def\cupress{Cambridge University Press}
\def\pup{Princeton University Press}
\def\wss{World Scientific, Singapore}
\def\oup{Oxford University Press}

\pagebreak
\pagestyle{empty}

\end{document}